\documentclass[11pt]{article}
\usepackage{amssymb}
\usepackage[dvips]{epsfig}
\usepackage{amsmath}

\begin{document}

\author{V. G. C. S. dos Santos, A. de Souza Dutra\thanks{%
E-mail: dutra@feg.unesp.br},  M. B. Hott\thanks{%
E-mail: hott@feg.unesp.br} \\
%EndAName
UNESP - Campus de Guaratinguet\'{a} - DFQ\\
12516-410 Guaratinguet\'{a} SP Brasil \\
}
\title{{\LARGE Real spectra for non-Hermitian Dirac equation in 1+1
dimensions with a most general coupling}}
\maketitle

\begin{abstract}
The most general combination of couplings of fermions with external
potentials in 1+1 dimensions, must include vector, scalar and pseudoscalar
potentials. We consider such a mixing of potentials in a PT-symmetric
time-independent Dirac equation. The Dirac equation is mapped into an
effective PT-symmetric Schr\"{o}dinger equation. Despite the non-hermiticity
of the effective potential, we find real energies for the fermion.

PACS numbers: 03.65.Ge, 03.65.Pm, 03.65.Fd
\end{abstract}

\section{Introduction}

In the last ten years the, so-called, PT symmetric systems introduced in the
seminal paper by Bender and Boettcher \cite{bender1} have attracted very
much attention. In fact, there are many works devoted to develop and
understand this new kind of situation \cite{jiadutra,mustafa}. They consist
of non-Hermitian Hamiltonians with real eigenvalues, which however exhibit
parity and time-reversal symmetries. The one-dimensional time-independent
Schr\"{o}dinger equation is invariant under space-time inversion. In
addition, there exist other classes of Hamiltonians with real spectra
without being PT symmetric, as can be seen, for instance, in the references
\cite{mostafazadeh,ahmed}; and systems where the PT symmetry is
spontaneously broken with complex energy eigenvalues \cite{bender2}. The
problem of non-Hermitian time-dependent interactions which does not exhibit
PT symmetry but still admits real energies is considered in the references
\cite{vah}-\cite{mostafazadeh2}.

More recently, some problems of relativistic fermions interacting with
non-Hermitian potentials of scalar and vector natures have been reported in
the literature \cite{jiadutra},\cite{mustafa}-\cite{mustafa2}. In general,
it has been shown that for some configurations of those non-Hermitian
potentials, the Dirac equation admits real energies. In 1+1 dimensions the
Dirac problem is easily mapped into a Sturm-Liouville problem or, in other
words, into a time-independent Sch\"{o}dinger equation with real or complex
potentials whose bound-state solutions present real energy eigenvalues. Some
authors \cite{9}-\cite{13} have investigated the Klein-Gordon equation and
Dirac equation in the context of PT symmetry and pseudo-Hermiticity. Mustafa
\cite{9} studied the exact energies for Klein-Gordon particle and Dirac
particle in the generalized complex Coulomb potential. Znojil \cite{10}
analyzed the Klein-Gordon equation presenting a pseudo-Hermiticity. Egrifes
et al \cite{11} investigated the bound states of the Klein-Gordon and Dirac
equations for the one-dimensional generalized Hulth\'{e}n potential within
the framework of PT-symmetric quantum mechanics. In \cite{12}, the authors
have investigated the bound states for the PT-symmetric versions of the
Rosen-Morse and Scarf II potentials in the Klein-Gordon equation with
equally mixed vector and scalar potentials. Sinha and Roy \cite{13}
investigated the one-dimensional solvable Dirac equation with non-Hermitian
scalar and pseudoscalar couplings, possessing real energy spectra.

The coupling with scalar potentials in the Klein-Gordon and Dirac equations
can be seen as a position-dependent effective mass. In the relativistic
ambiance, the ordering ambiguity of the mass and momentum operators, which
is present in the non-relativistic one, should disappear. Nevertheless,
there are difficulties to define consistently fermions and bosons, whenever
one takes into account space-time dependent masses. This happens due to the
fact that physical particles in quantum field theory must belong to an
irreducible representation of the Poincar\`{e} algebra \cite{14,15}. One
should be able to find generators specifying the particle properties,
usually its mass and helicity. However, it is quite hard to accomplish with
this task in the case of spatially dependent masses. Thus, one should keep
in mind that all of these usually thought as relativistic equations for
position-dependent masses should be taken as effective equations. In this
regard, Alhaidari \cite{17} studied the exact solution of the
three-dimensional Dirac equation for a charged particle with spherically
symmetric singular mass distribution in the Coulomb field. Vakarchuk \cite%
{18} investigated the exact solution of the Dirac equation for a particle
whose potential energy and mass are inversely proportional to the distance
from the force center. In \cite{peng}, the authors considered the smooth
step mass distribution and solved approximately the one-dimensional Dirac
equation with the spatially dependent mass for the generalized Hulth\'{e}n
potential. In \cite{20}, the authors investigated the exact solution of the
one-dimensional Klein-Gordon equation with the spatially dependent mass for
the inversely linear potential. Therefore, it is of considerable interest to
investigate the solution of the effective mass Klein-Gordon and Dirac
equations with non-Hermitian complex potentials with real energy spectra.

One interesting problem that has been tackled in this context is the Dirac
equation in 1+1 dimensions in the presence of a convenient complex vector
potential plus a real scalar potential \cite{jiadutra}, wherein the scalar
potential plays the role of a position dependent mass. Here, we will show
that one can realize a more general system of massless fermions in two
dimensions interacting with a mixing of complex vector, scalar and
pseudoscalar potentials. Although complex, the scalar and pseudoscalar
potentials are responsible to open a mass gap for the fermions. Such a
scenario might be important for some condensed matter systems, where the
electrical conduction is essentially one-dimensional. The scalar and
pseudoscalar potentials can be thought as defects in the lattice and the
electrons are also subject to a background potential of vector nature due to
the ions in the lattice.

Our purpose in this work is to show that the Dirac equation in a
two-dimensional world can still have real discrete energy spectrum and
supports fermion bound-states when a convenient mixing of complex vector,
scalar and pseudoscalar potentials is considered. We call attention to the
transformation of the potentials under parity in order to have the
PT-symmetry in the Dirac equation. Particular configurations of those
potentials are worked out in some detail. The approach here is the mapping
of the PT-symmetric Dirac problem into a naturally PT-symmetric effective
Schr\"{o}dinger equation.

\section{The time-independent Dirac equation in 1+1 dimensions}

We consider here the 1+1 dimensional time-independent Dirac equation for a
massless fermion under the action of a general potential $\mathcal{V}$. It
is written as%
\begin{equation}
H\Psi (x)=E\Psi (x)\,,  \label{eq1}
\end{equation}%
\begin{equation}
H=c\,\alpha \,p+\mathcal{V}\,,  \label{eq1a}
\end{equation}%
where $E$ is the energy of the fermion, $c$ is the velocity of light and $p$
is the momentum operator. $\alpha $ and $\beta $ are Hermitian square
matrices satisfying the relations $\alpha ^{2}=\beta ^{2}=1$, $\left\{
\alpha ,\beta \right\} =0$. The positive definite function $|\Psi |^{2}=\Psi
^{\dagger }\Psi $, satisfying a continuity equation, is interpreted as a
position probability density. This interpretation is completely satisfactory
for single-particle states.

We set $\mathcal{V}$ to be
\begin{equation}
\mathcal{V}=\beta \,M(x)+\beta \,\gamma _{5}\,P(x)+V(x)+\beta \,\alpha
\mathcal{A}(x)  \label{eq2}
\end{equation}%
where $M(x)$ is a scalar potential, $P(x)$ a pseudoscalar potential and $%
V(x) $ is the time-component of a Lorentzian 2-vector potential, whose space
component is $\mathcal{A}(x)$. The space component of the 2-vector potential
can be eliminated by a gauge transformation without affecting the physics.
Once we have only four linearly independent $2\times 2$ matrices, the
structure of coupling in $\mathcal{V}$ is the most general one can consider
in the time-independent Dirac equation in one space dimension.

In terms of the potentials the Hamiltonian (\ref{eq1a}) becomes
\begin{equation}
H=c\,\alpha \,p+V(x)+\beta \,M(x)+\beta \,\gamma ^{5}\,P(x)\,,  \label{eq2a}
\end{equation}%
\noindent where $\gamma ^{5}=-i\alpha $. An explicit expression for the $%
\alpha $ and $\beta $ matrices can be chosen from the Pauli matrices that
satisfy the same algebra. We use $\beta =\sigma _{1}$, $\alpha =\sigma _{3}$%
, and thus $\beta \gamma ^{5}=-\sigma _{2}$. The equation (\ref{eq1}) can be
decomposed in two coupled first-order differential equations, for the upper,
$\psi _{+}(x)$, and lower, $\psi _{-}(x)$, components of the spinor $\Psi
(x) $. In a simplified notation and by using the natural system of units $%
\hbar =c=1$, we have%
\begin{equation*}
-i\psi _{+}^{\prime }+M\psi _{-}+V\psi _{+}+iP\psi _{-}=E\psi _{+}~,
\end{equation*}%
\begin{equation}
+i\psi _{-}^{\prime }+M\psi _{+}+V\psi _{-}-iP\psi _{+}=E\psi _{-}~.
\label{eq3}
\end{equation}%
where the prime stands for the derivative with respect to $x$.

In the case that all the potentials are real functions, the Dirac equation
is Hermitian and invariant under space-reversal (parity) transformation. We
recall that the parity transformation is an improper Lorentz transformation
and that the spinor in one frame is constructed from the spinor in the other
frame by means of the relation $\tilde{\Psi}(\tilde{x},t)=S\Psi
(x,t)=e^{i\delta }\beta \Psi (x,t)$, with $\tilde{x}=-x\,\ $and $\delta $ an
overall constant phase factor. Moreover, under parity transformation, $M(x)$
and $V(x)$ do not change and $P(x)$ changes its sign. The matrices must
transform as $S^{-1}\beta S=\beta $ and $S^{-1}\alpha S=\alpha $.

The same transformations could be used even in the case that the potentials
are complex, but if we want the Dirac equation invariant under the
combination of parity and time-reversal transformations this issue becomes
trickier with a non-Hermitian Hamiltonian. This is because the time-reversal
transformation implies that $\mathcal{T}(i)\mathcal{T}^{-1}=-i$ and that the
potentials in the Hamiltonian are complex. Thus, althougt the time-reversal
does not change each part of the potentials, since they are
time-independent, it changes the relative sign between the real and the
imaginary parts of the potentials; as a consequence, the imaginary part of
each one of the potentials must change under parity in the reversed form of
its real part, in order to have the Dirac equation invariant under the
combination of parity and time-reversal transformations. In summary, in
order to have PT-symmetry even when the potentials are complex, the
imaginary part of the vector and scalar potentials must change their signs
under parity, whereas the imaginary part of the pseudoscalar potential does
not change. The spinor in the time-reversed system is obtained from the
spinor $\tilde{\Psi}(\tilde{x},t)$ by means of the following the
transformation $\tilde{\Psi}_{\mathcal{T}}(\tilde{x},\tilde{t})=\mathcal{T}%
\tilde{\Psi}(\tilde{x},t)=T\tilde{\Psi}^{\ast }(\tilde{x},t)$, with $\tilde{t%
}=-t$ and $T$ a square matrix such that $T^{-1}\beta ^{\ast }T=\beta $ and $%
T^{-1}\alpha ^{\ast }T=-\alpha $. Then $T$ must commute with $\beta $ and
anti-commute with $\alpha $, that is $T\equiv \beta $.

All the examples of potentials we are going to deal with follow the
transformation rule given above.

\section{The effective PT symmetric problem}

Whenever one considers only the coupling either with the scalar or the
pseudoscalar potential, the differential equations can be uncoupled in such
a way that both components of the spinor satisfy second-order differential
equations, similar to each other and to the Schr\"{o}dinger equation. By
including the coupling with the vector potential this is no longer possible.
Although, it is possible to show that one of the components obeys a kind of
Schr\"{o}dinger equation, and the other component is given in terms of the
previous one, as was done in the references in \cite{jiadutra}. From now on,
we are going to follow that approach. By applying the space derivative in
the first of the equations\ (\ref{eq3}) we have
\begin{equation}
-i\psi _{+}^{\prime \prime }+A_{+}\psi _{-}^{\prime }+A_{+}^{\prime }\psi
_{-}=B\psi _{+}^{\prime }+B^{\prime }\psi _{+}
\end{equation}%
where we have defined $A_{\pm }\equiv A_{\pm }(x)=M(x)\pm iP(x)$ and $%
B\equiv B(x)=E-V(x)$. By substituting\ in the above equation the expressions
for $\psi _{-}^{\prime }$ and $\psi _{-}$ taken from equations \ (\ref{eq3}%
), we obtain the following equation for the upper component
\begin{equation}
-i\psi _{+}^{\prime \prime }+i\frac{A_{+}^{\prime }}{A_{+}}\psi _{+}^{\prime
}+\left[ \frac{BA_{+}^{\prime }}{A_{+}}-B^{\prime }+i(A_{+}A_{-}-B^{2})%
\right] \psi _{+}=0,  \label{eq4}
\end{equation}%
and the equation obeyed by the lower component can be rewritten as
\begin{equation}
\psi _{-}=\frac{1}{A_{+}}\left( i\psi _{+}^{\prime }+B\psi _{+}\right) .
\label{eq5}
\end{equation}%
\noindent We notice that by means of the redefinition of the upper component
\begin{equation}
\psi _{+}(x)\equiv \sqrt{A_{+}}\chi (x)~,  \label{eq6}
\end{equation}%
the first-derivative term can be eliminated and $\chi $ satisfies the
differential equation
\begin{equation}
-i\chi ^{\prime \prime }+\left\{ \frac{BA_{+}^{\prime }}{A_{+}}-B^{\prime
}+i\left( A_{+}A_{-}-B^{2}\right) +i\left[ \frac{3}{4}\left( \frac{%
A_{+}^{\prime }}{A_{+}}\right) ^{2}-\frac{1}{2}\frac{A_{+}^{\prime \prime }}{%
A_{+}}\right] \right\} \chi =0,  \label{eq6a}
\end{equation}%
which, in its turn, can be written as a Schr\"{o}dinger, equation
\begin{equation}
-\chi ^{\prime \prime }+V_{eff}\chi =E^{2}\chi ,  \label{eq7}
\end{equation}%
where
\begin{equation}
V_{eff}=E\left( 2V+i\frac{A_{+}^{\prime }}{A_{+}}\right) -iV^{\prime
}+i\left( \frac{A_{+}^{\prime }}{A_{+}}\right) V+\left[ A_{+}A_{-}-V^{2}+%
\frac{3}{4}\left( \frac{A_{+}^{\prime }}{A_{+}}\right) ^{2}-\frac{1}{2}%
\left( \frac{A_{+}^{\prime \prime }}{A_{+}}\right) \right] .  \label{eq8}
\end{equation}

Equations (\ref{eq4})-(\ref{eq8}) constitute the essential set of equations
we are going to deal with throughout this work. Our intention is to find
convenient configurations of the potentials for which the equation \ (\ref%
{eq7}) presents normalizable solutions and positive eigenvalues or, in other
words, bound-state solutions corresponding to a discrete energy spectrum,
with $E^{2}\geq 0$. Naturally, the solutions for the lower component that
satisfies equation \ (\ref{eq5}) must also be normalizable.

In the literature we find several studies of the Dirac equation in a
two-dimensional world with convenient mixing of scalar plus vector
potentials and pseudoscalar plus vector potentials. As far as we know, most
of them assume an intrinsic relation between the scalar and vector
potential, or between the pseudoscalar and the vector potential. A typical
case is the one which deals with a constraint of the type $M(x)=\kappa V(x)$%
, where $|\kappa |>0$ and $M(x)$ is a binding potential \cite{castrohott1}.
There are some cases in four dimensions for which such convenient relation
leads to analytical and exact solutions for the Dirac equation. Here, we are
not going to escape from a constraint among the potentials. On the contrary,
we follow the same procedure carried out in the references \cite{jiadutra}
by constraining the vector potential to obey the more general relation
\begin{equation}
V=\frac{i}{2}\left( \frac{A_{+}^{\prime }}{A_{+}}\right) +\widetilde{V}=%
\frac{i}{2}\left( \frac{M^{\prime }+iP^{\prime }}{M+iP}\right) +\widetilde{V}%
.  \label{eq9}
\end{equation}%
With this relation the Schr\"{o}dinger-like equation (\ref{eq7}) is given by
\begin{equation}
-\chi ^{\prime \prime }+\left[ M^{2}+P^{2}-\widetilde{V}^{2}+2E\widetilde{V}%
~-i\widetilde{V}^{\prime }\right] \chi =E^{2}\chi ~.  \label{eq10}
\end{equation}

The effective potential written explicitly in terms of the real and
imaginary parts of the potentials

\begin{eqnarray}
V_{eff}(x) &=&\left( M_{r}^{2}-M_{i}^{2}+P_{r}^{2}-P_{i}^{2}-\widetilde{V_{r}%
}^{2}+\widetilde{V_{i}}^{2}+2E\widetilde{V}_{r}~+\widetilde{V_{i}}^{\prime
}\right) +  \notag \\
&&+2i\left( M_{r}M_{i}+P_{r}P_{i}-\tilde{V}_{r}\tilde{V}_{i}+2E\tilde{V}_{i}-%
\tilde{V}_{r}^{\prime }\right)  \label{vef}
\end{eqnarray}%
is complex, but the effective Schr\"{o}dinger equation exhibits PT symmetry
if $M_{r}$, $\tilde{V}_{r}$ and $P_{i}$ remains invariant, whereas $M_{i}$, $%
\tilde{V}_{i}$ and $P_{r}$ changes their signs under parity transformation,
as was stated in the second section. Naturally we are thinking of the cases
for which the energy in the Dirac problem is real (positive or negative).

\section{Real spectrum with non-Hermitian potentials}

\ As our first example, we choose a set of potentials which have a quadratic
dependence on the spatial variable and, in order to deal with a complete set
of exact eigenstates, we impose some restrictions over the potential
parameters. In this first example we choose%
\begin{equation}
M(x)=a_{0}x^{2},~P(x)=i(a_{1}x^{2}+b_{1}),~\tilde{V}(x)=a_{2}x^{2}
\end{equation}%
In this situation, the vector potential is given by

\begin{equation}
V=i\left[ \frac{(a_{0}-a_{1})x}{a_{0}x^{2}-(a_{1}x^{2}+b1)}\right]
+a_{2}x^{2}.
\end{equation}%
Consequently the effective potential is

\begin{equation}
Veff=(a_{0}^{2}-a_{2}^{2}-a_{1}^{2})x^{4}+(2Ea_{2}-2a_{1}b_{1})x^{2}-b_{1}^{2}-2ia_{2}x.
\end{equation}

\noindent As we asserted, we want to deal with an exactly solvable model. We
restrict the potential parameters in order to eliminate the quartic term,
obtaining the constraint $a_{0}=\sqrt{a_{1}^{2}+a_{2}^{2}}$. This lead us to
the following effective PT-symmetric driven harmonic potential

\begin{equation}
Veff=\Omega ^{2}x^{2}-2ia_{2}x-b_{1}^{2}  \label{eq. Veff3}
\end{equation}

\noindent where $\Omega ^{2}=2Ea_{2}-2a_{1}b_{1}$. \ By means of the
transformation $x=y+2ia_{2}/\Omega ^{2}$ we obtain the following
differential equation for $\chi (y)$

\begin{equation}
-\frac{d^{2}\chi }{dy^{2}}+\Omega ^{2}y^{2}\chi =\left[ E^{2}+b_{1}^{2}-%
\frac{a_{2}^{2}}{\Omega (E)^{2}}\right] \chi ,
\end{equation}%
that is a quantum harmonic oscillator of \ unity mass. Consequently the
energy $E$ of the relativistic system is obtatined from the following
equation

\begin{equation}
E^{2}=-b_{1}^{2}+\frac{a_{2}^{2}}{\Omega \left( E\right) ^{2}}+(2n+1)~\Omega
\left( E\right) .  \label{especeq}
\end{equation}

\noindent The above equation is considerably simplified if we set $b_{1}=0$,
that is%
\begin{equation*}
E^{2}=\frac{a_{2}}{2~E}+\left( 2~n+1\right) \sqrt{2~E~a_{2}},
\end{equation*}%
which can be written as%
\begin{equation*}
\left( E^{3}-\frac{a_{2}}{2}\right) =\left( 2~n+1\right) E\sqrt{2~E~a_{2}},
\end{equation*}%
and, once squared, lead us to%
\begin{equation*}
e^{2}-a_{2}\left[ 1+2\left( 2~n+1\right) ^{2}\right] ~e+\frac{a_{2}^{2}}{4}%
=0,
\end{equation*}%
where we have defined $e\equiv E^{3}$, leading finally to the solutions%
\begin{equation*}
E_{\pm }=\left( \frac{a_{2}}{2}\left\{ 1+2\left( 2~n+1\right) ^{2}\pm \sqrt{%
\left[ 1+2\left( 2~n+1\right) ^{2}\right] ^{2}-1}\right\} \right) ^{\frac{1}{%
3}}.
\end{equation*}

\bigskip

One can note that it is not necessary the presence of $\tilde{V}(x)$ in
order to have an effective PT-symmetric Schr\"{o}dinger equation with a
complex effective potential. As a matter of fact one can see that the
dependence of $\tilde{V}$ on the coordinate implies in an effective
potential explicitly dependent on the energy of the Dirac problem and, as a
consequence one will reach invariably an intricate transcendental equation
for the energy. Henceforth, we set $\tilde{V}$ a real constant.

As an example, we suggest the following configurations for the scalar and
the pseudoscalar potentials%
\begin{equation}
M(x)=i\omega _{1}x+m_{1},~P(x)=\omega _{2}x+im_{2}  \label{eq10a}
\end{equation}%
with $\omega _{1}$, $m_{1}$, $\omega _{2}$ and $m_{2}$ real constants. The
vector potential, with $\widetilde{V}=0$, is complex and given by%
\begin{equation}
V(x)=-\frac{(\omega _{1}-\omega _{2})\left[ (m_{1}-m_{2})+i(\omega
_{1}+\omega _{2})x\right] }{2\left[ (\omega _{1}+\omega
_{2})^{2}x^{2}+(m_{1}-m_{2})^{2}\right] },  \label{eq11}
\end{equation}%
and the effective potential in the Schr\"{o}dinger equation obeyed by $\chi
(x)$

\begin{equation}
V_{eff}(x)=\omega ^{2}x^{2}+2i(\omega _{1}m_{1}+\omega
_{2}m_{2})x+(m_{1}^{2}-m_{2}^{2}),  \label{eq12}
\end{equation}%
with $\omega ^{2}=$ $\omega _{2}^{2}-\omega _{1}^{2}$, is PT-symmetric.

By defining the variable $y=x-i(\omega _{1}m_{1}+\omega _{2}m_{2})/\omega
^{2}$ we obtain%
\begin{equation}
-\frac{d^{2}\chi }{dy^{2}}+\omega ^{2}\left( y^{2}+\lambda ^{2}\right) \chi
=E^{2}\chi ~,  \label{eq13}
\end{equation}%
whose non-normalized eigenfunctions and the eigenvalues are respectively
given by
\begin{equation}
\chi _{n}(y)=e^{-\omega y^{2}/2}H_{n}(\sqrt{\omega }y)  \label{eq14}
\end{equation}%
and%
\begin{equation}
E^{2}=\omega (2n+1)+\lambda ^{2}\omega ^{2},  \label{eq15}
\end{equation}%
where $H_{n}(\sqrt{\omega }y)$ is the Hermite polynomial of degree $n$ and $%
\lambda ^{2}=\left( \frac{\omega _{1}m_{2}+\omega _{2}m_{1}}{\omega ^{2}}%
\right) ^{2}$. Consequently, the upper component of the Dirac spinor can be
written as%
\begin{equation}
\psi _{+,n}(y)=N\sqrt{\omega }(y^{2}+\lambda ^{2})^{1/4}e^{-\frac{1}{2}%
(\omega y^{2}-i\theta (y))}H_{n}(\sqrt{\omega }y),  \label{eq16}
\end{equation}%
where $\theta (y)=\tan ^{-1}(y/\lambda )$, and $N$ is a normalization
constant. By recalling the equation (\ref{eq5}) and the expression for the
vector potential, $V(x)=iA_{+}^{\prime }/2A_{+}$, we get the following
equation for the lower component of the spinor%
\begin{equation}
\psi _{-}(y)=\frac{1}{\sqrt{A_{+}(y)}}\left( i\frac{d\chi }{dy}+E\chi
\right) ,  \label{eq17}
\end{equation}%
which can be written in terms of Hermite polynomials as%
\begin{equation}
\psi _{-,n}(y)=\frac{N}{\sqrt{\omega }(y^{2}+\lambda ^{2})^{1/4}}e^{-\frac{1%
}{2}(\omega y^{2}+i\theta (y))}\left[ \left( \pm \left\vert E_{n}\right\vert
-i\omega y\right) H_{n}(\sqrt{\omega }y)+i2n\sqrt{\omega }H_{n-1}(\sqrt{%
\omega }y)\right] .  \label{eq18}
\end{equation}%
The upper (lower) sign of the factor $\pm \left\vert E_{n}\right\vert $ in
the expression above stands for the positive (negative) energy solution and $%
\left\vert E_{n}\right\vert =\sqrt{\omega (2n+1)+\lambda ^{2}\omega ^{2}}$.
The normalization constant $N$ \ is obtained by the definition of the norm
of the eigenstates
\begin{equation}
\int_{-\infty }^{+\infty }|\Psi _{n}|^{2}dx=\int_{-\infty }^{+\infty }(|\psi
_{+,n}(y)|^{2}+|\psi _{-,n}(y)|)^{2}dy=1\text{,}  \label{eq19}
\end{equation}%
which can be written in terms of the variable $u=\sqrt{\omega }y$ as%
\begin{equation}
2\left\vert N\right\vert ^{2}\int_{0}^{+\infty }du~\frac{e^{-u^{2}}}{%
(u^{2}+\lambda ^{2}\omega )^{1/2}}\left\{ \left[ 2(u^{2}+\lambda ^{2}\omega
)+(2n+1)\right] H_{n}^{2}(u)-2nH_{n-1}(u)H_{n+1}(u)\right\} =1,  \label{eq20}
\end{equation}%
where we have used the recursion relation $%
2uH_{n}(u)=2nH_{n-1}(u)+H_{n+1}(u) $.~It is very hard to check the
convergence of the above integral, for all values of $n$, even though each
of them converges, since the weight function decays faster, for $%
x\rightarrow \pm \infty $, than any polynomial appearing in the numerator of
the integrand. We furnish below the normalization constant for the two
lowest energy levels,
\begin{eqnarray*}
N_{0} &=&\left\{ e^{-\lambda ^{2}\omega /2}(1+2\lambda ^{2}\omega
)K_{0}(\lambda ^{2}\omega /2)+\sqrt{\pi }U(1/2,0,\lambda ^{2}\omega
)\right\} ^{-1/2}, \\
N_{1} &=&\left\{ e^{-\lambda ^{2}\omega /2}\left[ 2(1+\lambda ^{4}\omega
^{2})K_{0}(\lambda ^{2}\omega /2)+2\lambda ^{2}\omega (1-\lambda ^{2}\omega
)K_{1}(\lambda ^{2}\omega /2)\right] +\right. \\
&&\left. +\sqrt{\pi }(1+2\lambda ^{2}\omega )U(1/2,0,\lambda ^{2}\omega
)\right\} ^{-1/2},
\end{eqnarray*}%
where $K_{0}(z/2)$ and $K_{1}(z/2)$ are modified Bessel functions of the
second kind and $U(a,b,z)$ is the confluent hypergeometric function.

As a third example we take the scalar and pseudoscalar potentials given
respectively by
\begin{equation}
M=iM_{1}\tanh (\mu x)+M_{0},~P=P_{1}\tanh (\mu x)+iP_{0},  \label{eq21}
\end{equation}%
with $M_{1}$, $P_{1}$, $M_{0}$, $P_{0}$ reals and constants. Thus, the
vector potential is also complex and, with $\tilde{V}=V_{0}$, is given by,%
\begin{equation}
V=-\frac{\mu }{2}\frac{(P_{1}+M_{1})~\left[ (M_{0}-P_{0})-i(M_{1}+P_{1})%
\tanh (\mu x)\right] \mathrm{sech}(\mu x)}{%
(M_{0}-P_{0})^{2}+(M_{1}+P_{1})^{2}\tanh ^{2}(\mu x)}+V_{0},  \label{eq21a}
\end{equation}%
and the effective Schr\"{o}dinger equation (\ref{eq10}) can be written as
\begin{equation}
-\chi ^{\prime \prime }-V_{eff}\chi =\varepsilon \chi ~,  \label{eq22}
\end{equation}%
where
\begin{equation}
V_{eff}(x)=-(P_{1}^{2}-M_{1}^{2})\mathrm{sech}^{2}(\mu
x)+2i(M_{0}M_{1}+P_{0}P_{1})\tanh (\mu x)  \label{eq22a}
\end{equation}%
and
\begin{equation}
\varepsilon =(E-V_{0})^{2}-(P_{1}^{2}-M_{1}^{2})+(P_{0}^{2}-M_{0}^{2}).
\label{eq22b}
\end{equation}%
This effective potential is a generalization of the real Rosen-Morse II
(hyperbolic) \cite{rosen, khare} and was already considered in the
literature concerning PT symmetry in non-Hermitian Schr\"{o}dinger problems
\cite{mostafazadeh},\cite{ahmed}. The eigenvalues of the effective energy
can be conveniently written as%
\begin{equation}
\varepsilon _{n}=-\mu ^{2}(a_{n}^{2}-b_{n}^{2}),  \label{eq23}
\end{equation}%
where%
\begin{equation}
a_{n}=(1/2)\left[ \sqrt{1+\left( 2U_{1}/\mu \right) ^{2}}-(2n+1)\right]
\text{ \ \textrm{and} \ }b_{n}=\Omega /\mu ^{2}a_{n},  \label{eq24}
\end{equation}%
with $U_{1}^{2}=P_{1}^{2}-M_{1}^{2}$, $\Omega =M_{0}M_{1}+P_{0}P_{1}$ and $%
n=n=0,1,2,...<$ $s=(1/2)\left[ \sqrt{1+\left( 2U_{1}/\mu \right) ^{2}}-1%
\right] $. Thus, the energy eigenvalues of the Dirac problem are given by
the following relation%
\begin{equation}
E_{n}=V_{0}\pm \sqrt{U_{1}^{2}-U_{0}^{2}-\mu ^{2}(a_{n}^{2}-b_{n}^{2})},
\label{eq25}
\end{equation}%
where $U_{0}^{2}=P_{0}^{2}-M_{0}^{2}$. In order to have a real spectra, the
relation\ $U_{1}^{2}-U_{0}^{2}\geq \mu ^{2}(a_{n}^{2}-b_{n}^{2})$ among the
parameters of the potential must be imposed. It worths mentioning that this
kind of restriction is not a consequence of a complex potential in the
effective Schr\"{o}dinger equation. It appears, instead, due to the kind of
coupling and to the configuration of the potentials we have considered in
the relativistic problem.

If the potential supports $n_{\max }+1$ bound states, then we can say that $%
s=n_{\max }+\delta $, with $0<\delta <1$. We have observed that, for $\Omega
^{2}=\mu ^{4}\delta ^{4}$, the factor $\mu ^{2}(a_{n}^{2}-b_{n}^{2})$
decreases in the range $\left[ 0,n_{\max }\right] $. The maximum value
achieved by $\mu ^{2}(a_{n}^{2}-b_{n}^{2})$ is $\mu ^{2}(s^{2}-\delta
^{4}/s^{2})\,$. Therefore, the relations%
\begin{equation}
\Omega ^{2}=\mu ^{4}\delta ^{4},~U_{1}^{2}-U_{0}^{2}>\mu ^{2}\left[ (n_{\max
}+\delta )^{2}-\frac{\delta ^{4}}{(n_{\max }+\delta )^{2}}\right]
\label{eq26}
\end{equation}%
are necessary to have a real spectrum. In the case we have only one bound
state for example, corresponding to $n_{\max }=0$, we obtain $%
M_{0}M_{1}+P_{0}P_{1}=\mu ^{2}\delta ^{2}$, $%
P_{0}^{2}-M_{0}^{2}<P_{1}^{2}-M_{1}^{2}=\mu ^{2}\delta (\delta +1)$.

In general the upper component of the spinor is given by

\begin{equation}
\psi _{+,n}(z)=N_{+}|A_{+}(z)|^{1/2}e^{i\alpha
(z)/2}(1-z^{2})^{a_{n}/2}\left( \frac{1-z}{1+z}\right)
^{b_{n}}P_{n}^{(a_{n}+b_{n},a_{n}-b_{n})}(z),  \label{eq30}
\end{equation}%
where, $z=\tanh (\mu x)$, $P_{n}^{(\alpha ,\beta )}(z)$ are the Jacobi
polynomials of degree $n$ \cite{abram}.\ On its turn the lower component can
be written as%
\begin{eqnarray}
\psi _{-,n}(z) &=&\left. N_{-}|A_{+}(z)|^{-1/2}e^{-i\alpha (z)/2}\mu
(1-z^{2})^{a_{n}/2}\left( \frac{1-z}{1+z}\right) ^{b_{n}}\times \right.
\notag \\
&&\left. \times \left[
(E_{n}-a_{n}z-2b_{n}-V_{0})P_{n}^{(a_{n}+b_{n},a_{n}-b_{n})}+(1-z^{2})\frac{d%
}{dz}P_{n}^{(a_{n}+b_{n},a_{n}-b_{n})}\right] \right. .  \label{eq31}
\end{eqnarray}%
In the expressions\ (\ref{eq30}) and (\ref{eq31}) we have $|A_{+}(z)|=\sqrt{%
(M_{0}-P_{1}z)^{2}+(M_{1}z+P_{0})^{2}}$ and $\alpha (z)=\tan
^{-1}((M_{1}z+P_{0})/(M_{0}-P_{1}z))$.

The normalization constants are found by the definition of the norm, namely%
\begin{equation*}
\int_{-1}^{+1}dz~(1-z^{2})^{-1}(\left\vert \psi _{+,n}\right\vert
^{2}+\left\vert \psi _{-,n}\right\vert ^{2})=1.
\end{equation*}

\section{Final considerations}

We have extended the problem of fermions in 1+1 dimensions interacting with
complex potentials by considering the most general coupling of fermions
which is possible in the time-independent Dirac equation in two dimensions.
The potentials are considered to be of scalar, pseudoscalar and vector
natures. We have shown that one can map the Dirac equation with all those
complex potentials into a PT-symmetric effective Schr\"{o}dinger equation,
which can be exactly solved and present real spectra. Indeed, the
PT-symmetry is also observed already in the Dirac equation when the
imaginary part of each one of the potentials transforms, under parity
transformation, in the reversed manner that the real part of the potential.
In this way, the PT-symmetry seems to play a fundamental role for the
reality of the spectrum.\ There still remains the open question about
pseudo-hermiticity of the Dirac operator with the mixing of potentials we
have considered.

Another remarkable point of those examples, besides being exactly solvable,
is that they bind fermions despite the presence of the vector potential.
Inhomogeneous vector potentials give rise to electric fields which are
responsible for pair production. Based on what happens when we have a
convenient mixing of scalar and vector potential \cite{castrohott1}, we
think that the presence of the scalar and pseudoscalar potentials, although
complex, are responsible for giving an effective mass to the fermion and, as
a consequence, the threshold for pair production becomes higher than the
energy that the electric field can supplies to the fermion. Confinement of
fermions with a vector potential, as in the second example treated here, is
a peculiar case. Indeed, one can check that the relativistic spectra we have
found is very similar to that of the \textquotedblleft Dirac
oscillator\textquotedblright\ in 1+1 dimensions \cite{diracosc1}. In that
reference a massive fermion is coupled to a linear potential. That coupling,
being of a pseudoscalar type, is the two-dimensional version of the
four-dimensional coupling of an anomalous magnetic moment with an electric
field which emerges in the \textquotedblleft Dirac
oscillator\textquotedblright\ \cite{moshinsky}-\cite{mo2}, and that can be
seen as a mechanism for confinement of fermions in 3+1 dimensions.

We have noted that in the absence of the scalar and pseudoscalar couplings,
the effective equation for the upper component of the spinor is given by

\begin{equation*}
-\psi _{+}^{\prime \prime }+\left[ 2~i~V^{\prime }-(E-V)^{2})\right] \psi
_{+}=0,
\end{equation*}%
that is, we have a PT-symmetric effective Schr\"{o}dinger equation with a
vector potential which is real and even under parity transformation. This is
a situation where the PT-symmetry becomes quite important, because it opens
the possibility of having real spectra in a model which, otherwise, would be
discarded due to the fact that the effective Schr\"{o}dinger equation
corresponds to a non-Hermitian operator. This opens the promising
perspective of finding exactly solvable real potentials where this feature
could be accomplished.

\bigskip

\noindent\textbf{Acknowledgments: }

This work was supported in part by means of funds provided by CAPES, CNPq
and FAPESP. This work was partially done during a visit (ASD) within the
Associate Scheme of the Abdus Salam ICTP.

\end{document}